\documentclass[prl,aps,twocolumn,superscriptaddress]{revtex4-1}
\usepackage{graphicx,color}
\usepackage{amsthm}
\usepackage{amsfonts}
\usepackage{algorithmic}
\usepackage{enumerate}
\usepackage{latexsym}
\usepackage{amsmath}
\usepackage{amssymb}
\usepackage[colorlinks=true,citecolor=blue,linkcolor=blue]{hyperref}

\emergencystretch=\maxdimen
\begin{document}

\title{Mott insulating state and $d+id$ superconductivity in an ABC graphene trilayer}

\author{Huijia Dai}
\affiliation{Department of Physics, Beijing Normal University, Beijing 100875, China\\}
\author{Junchen Hou}
\affiliation{Department of Physics, Beijing Normal University, Beijing 100875, China\\}
\author{Xiao Zhang}
\affiliation{Department of Physics, Beijing Normal University, Beijing 100875, China\\}
\author{Ying Liang}
\affiliation{Department of Physics, Beijing Normal University, Beijing 100875, China\\}
\author{Tianxing Ma}
\email{txma@bnu.edu.cn}
\affiliation{Department of Physics, Beijing Normal University, Beijing 100875, China\\}
\affiliation{Beijing Computational Science Research Center, Beijing
100193, China}

\begin{abstract}
Motivated by the recently experimental reported signatures of the tunable Mott insulating state and superconductivity in an ABC graphene trilayer superlattice,
we investigate the charge compressibility, the spin correlation, and the superconducting instability
within the Hubbard model on a three-layer honeycomb lattice.
It is found that an antiferromagnetically ordered Mott insulator emerges beyond a critical $U_c$ at half-filling,
and the electronic correlation drives a $d+id$ superconducting pairing to be dominant over other pairing patterns in a wide doped region.
The effective pairing interaction with $d+id$ pairing symmetry is strongly enhanced with the increasing of on-site interaction, and suppressed as the interlayer coupling strength increases. Our intensive numerical results demonstrate that the insulating state and superconductivity in an ABC graphene trilayer are driven by strong electric correlation, and it may offer an attractive systems to explore rich correlated behaviors.
\end{abstract}

\noindent


\maketitle

\noindent
\underline{\it Introduction}---
Success in isolating atomically thin graphene systems has led to an explosion of interests in exploring their novel correlated electronic properties\cite{Novoselov2005Two,Zhang2005Experimental,RevModPhys.83.407}.
The well-known example is the magic-angle twisted bilayer graphene (TBG), a pair of stacked monolayer graphenes rotated in a particular angle in which the superconductivity and correlated insulating states have been observed experimentally\cite{Cao2018A,Cao2018B,Yankowitz1059}.
TBG captivates researchers due to their structural simplicity, and it offers a platform to explore the complex physics of superconductivity, which is a central problem in condensed
matter physics. Along with the progress in bilayer graphene, more and more attention turned to trilayer graphene (TLG), even multilayer graphene, which has more complex interlayer interactions and supplies a richer electronic structure.

Generally, there are two typical ways to stack the graphene layers, i.e., bernal stacking and rhombohedral stacking.
We refer to bernal stacked graphene as ABA graphene and rhombohedral stacked graphene as ABC graphene.
As the interlayer coupling strongly modifies the linear dispersion of monolayer graphene, the electronic structures are various in multilayer graphene films.
The ABA-TLG shows linear and parabolic dispersions, presenting a semimetallic property with a small band overlap at the Dirac point, and the ABC-TLG shows only parabolic dispersions, behaving such as a semiconductor as a band gap about 20 meV near the Dirac point\cite{PhysRevLett.97.036803,Aoki2007,APL_Kumar,JPCC_Tang}.
The band structure can also be changed by applying a perpendicular electric field.
Theoretical\cite{PhysRevB.81.125304,PhysRevB.82.035409,APL_Kumar,APL_Wu,JPCC_Tang} and experimental\cite{Lui2011,Bao2011} research have proved that the band gap of ABC-TLG is tunable with the external electric field, which is similar to the phenomenon reported in bilayer graphene\cite{PhysRevLett.99.216802,APL92_24,2009Direct,CRACIUN201142}.
Coincidentally, recent experiments discovered signs of the correlated insulating states\cite{NaturePhysics_Chen} and tunable superconductivity\cite{Nature572_Chen} in ABC graphene trilayer on hexagonal boron nitride (hBN).
Comparing to twisted bilayer graphene\cite{Cao2018A,Cao2018B}, ABC-TLG/hBN also exhibits the $moir\acute{e}$-induced physics such as the formation of the secondary Dirac bands and the miniband structure.
The combination of the energy dispersion in ABC-TLG and the narrow electronic minibands induced by the $moir\acute{e}$ potential leads to the observation of insulating states for the Mott insulator.
Evidence found in bilayers\cite{Cao2018A} and trilayers\cite{NaturePhysics_Chen} both show that the long-period $moir\acute{e}$ interference pattern significantly modifies the Dirac dispersion, and a correlated Mott insulating state occurs when such a miniband contains an integer number of electrons per superlattice unit cell.
Besides, gate tuning the charge density away from the half-filling, the Mott insulator led to superconductivity with strong coupling characteristics\cite{Cao2018B}.
These fascinating phenomena show a number of similarities with that of doped cuprates\cite{Bednorz1986} for which superconductivity occurs proximate to a Mott insulator.
The finding raises the intriguing possibility of graphene $moir\acute{e}$ superlattices serving as a new platform for studying unconventional superconductivity.

Motivated by the experimental discoveries, a great deal of theoretical efforts have been undertaken on the detailed properties of the possible nature of the exotic correlated electronic phases in the graphene superlattice\cite{Meng.Nature.464,14PhysRevB.98.045103,15PhysRevX.8.031089,16PhysRevLett.121.087001,20PhysRevX.8.041041,PhysRevB.99.075127,Chu_2020,PhysRevB.102.245105,PhysRevB.84.121410,Lin2015Quantum,PhysRevB.90.245114,Ma2015EPL}.
However, the mechanism of the superconductivity and the correlated insulating state in ABC-TLG are still under very active debate and a lot of works need to implement\cite{PhysRevLett.122.016401,PhysRevB.99.205150,PhysRevLett.124.187601,pantaleon2020narrow}.
The goal of the current paper is to understand the nature of superconducting phases and the correlated insulating state in ABC-TLG, specifically,
to identify the doping dependent dominant superconducting pairing symmetry and magnetic order at half-filling.

Considering the strong correlation effect dominates in the system,
the unbiased numerical techniques are believed to be the appropriate approach to reveal its rich correlated behavior.
We focus on the Mott physics and superconducting pairing correlation in ABC-TLG.
By using the determinant quantum Monte Carlo method, the behavior of
charge compressibility and spin correlation at half-filling is examined, which reveal an antiferromagnetically ordered Mott insulator emerges beyond a
critical $U_c$. Our simulation shows that the superconducting pairing correlation with the $d+id$ wave dominates over other pairing symmetries, which is similar with our previous studies\cite{PhysRevB.84.121410,HUANG2019310,ModernPhysicsLettersB.Zhang,PhysRevB.101.155413}, suggesting that dominant pairing symmetry of the superconductivity emerged in graphene systems is mainly determined by the inherent honeycomb lattice structure of graphite.
For further study, we considered the effect of on-site interaction $U$ and interlayer interaction $t_c$, and it is found that the superconducting pairing correlation with $d+id$ wave symmetry is readily enhanced by the existence of $U$ and slightly suppressed with $t_c$.
Our extensive numerical results verify the viewpoint of that superconductivity in ABC graphene trilayer arises in a doped Mott insulator.

\begin{figure}[tbp]
\includegraphics[scale=0.5]{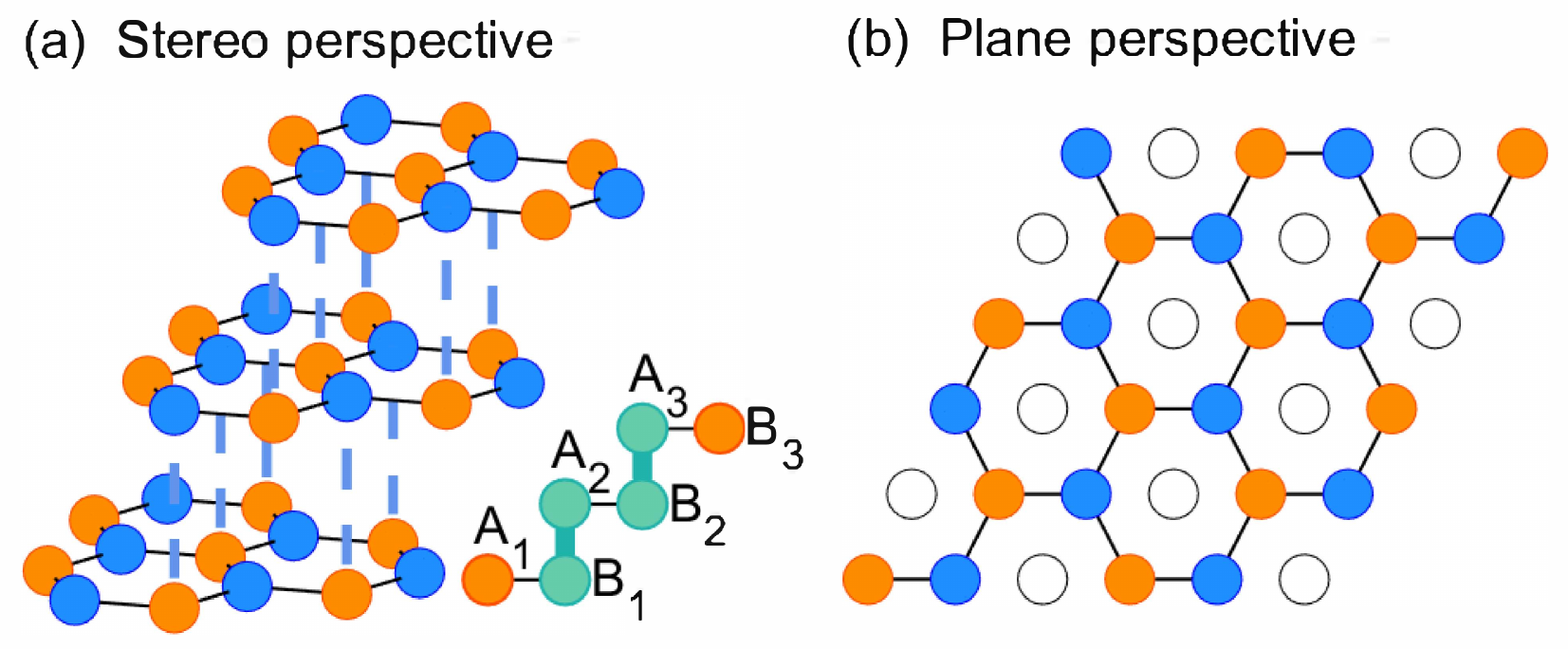}
\caption{(a) Three-dimensional crystal structure diagram and interlayer hopping processes for the graphene trilayer with ABC stacking order.
(b) Planar structure schematic of the honeycomb lattice with linear lattice size $L=6$.
The orange and blue full symbols represent the A and B sublattices of the honeycomb structure together with the empty symbols forming the underlying triangular lattice.
Here the designed honeycomb lattice has $2L^2/3$ sites, which is a $2/3$ subset of the triangular lattice with $L^2$ sites.
} \label{structure}
\end{figure}

\noindent
\underline{\it Model and Methods}
The structure of the ABC-stacked graphene trilayer is sketched in Fig.\ref{structure}.
In this geometry, each lattice consists of three layers which is staggered from each other, and each layer has interpenetrating triangular A and B sublattices.
Every adjacent layer pair forms an AB-stacked bilayer with the upper B sublattice directly on top of the lower A sublattice and the upper A above the center of a hexagonal plaquette of the layer below. Considering the electronic correlation, the Hubbard Hamiltonian reads
\begin{equation}
\begin{split}
H = &-t\sum_{\langle{ij}\rangle\sigma}\sum_{l=1}^{3}[a_{il\sigma}^{\dagger}b_{jl\sigma}+\text{H}.\text{c}.]\\
&-t_{\bot}\sum_{i\sigma}[b_{i1\sigma}^{\dagger}a_{i2\sigma}+b_{i2\sigma}^{\dagger}a_{i3\sigma}+\text{H}.\text{c}.]\\
&+\mu\sum_{i\sigma}\sum_{l=1}^{3}(a_{il\sigma}^{\dagger}a_{il\sigma}+b_{il\sigma}^{\dagger}b_{il\sigma})\\
&+U\sum_{i}\sum_{l=1}^{3}(n_{ila\uparrow}n_{ila\downarrow}+n_{ilb\uparrow}n_{ilb\downarrow}).
\end{split}
\label{1}
\end{equation}
Here $a_{il\sigma}^{\dagger}(a_{il\sigma})$ are annihilation (creation) operators which act at site ${\textbf{R}}_{li}^a$ of $l$ ($l=1,2,3$) layer with spin $\sigma(\sigma = \uparrow ,\downarrow)$ on sublattice A, and $b_{il\sigma}^{\dagger}(b_{il\sigma})$ acts similarly on sublattice B.
Occupy number operators $n_{ila\sigma} = a_{il\sigma}^{\dagger}a_{il\sigma}$ and $n_{ilb\sigma} = b_{il\sigma}^{\dagger}b_{il\sigma}$.
$t \approx 2.7eV$ denotes the in-plane hopping amplitude between nearest-neighbor $(NN)$, which is chosen to set the energy scale in the following, and $t_{\bot}$ denotes the interlayer hopping energy in the perpendicular direction to the NN bond.
$\mu$ and $U$ are the chemical potential and the on-site interaction strength.
The interlayer coupling energy $t_c=t_{\bot}$ is about $t_c=0.138t$, which is taken from that of Ref.\cite{Lui2011}.

Our simulations are mostly performed on a lattice of $L=9$ with periodic boundary conditions.
$L$ is the linear dimension of the lattice, which corresponds to the linear dimension of the underlying triangular lattice as shown in Fig.\ref{structure}(b).
It is a much more tough job to simulate ABC-TLG numerically than that of bilayer graphene systems.
The choice of the basic sketch in Fig.\ref{structure}(b) allows us to have the finite-size scaling for ABC-TLG,
and lattices with $L = 6,9,12,15$, and $18$ are simulated.
The number of lattice sites in each layer is $2L^2/3$\cite{PhysRevB.72.085123} where the number 2 means two inequivalent triangular sublattices, and the total number is $N_s = {3}\times2\times{L^2}/3$.
The basic strategy of the finite-temperature determinant quantum Monte Carlo method\cite{PhysRevD.24.2278} is to express the partition function as a path integral over the discretized inverse temperature over a set of random auxiliary fields.
The integral is then accomplished by Monte Carlo techniques.
In our simulations, 4000 sweeps were used to equilibrate the system, and an additional $12000-200000$ sweeps were then performed, each of which generated a measurement.
These measurements were divided into $20$ bins that provide the basis of coarse-grain averages, and errors were estimated based on standard deviations from the average.
In order to assess our results and their accuracy with respect to the infamous sign problem as the particle-hole symmetry is broken, a very careful analysis on the average of sign is shown.

To study the possible metal-insulator transition, we examine the $T$-dependent dc conductivity calculated from the wave-vector $\textbf{q}$ and the imaginary time $\tau$-dependent current-current correlation function $\Lambda_{xx}(\textbf{q},\tau)$,
\begin{equation}
\sigma_{dc}(T) = \frac{{\beta}^2}{\pi}\Lambda_{xx}(\textbf{q},\tau)(\textbf{q} = 0,\tau = \frac{\beta}{2}),
\label{2}
\end{equation}
where $\Lambda_{xx}(\textbf{q},\tau) = \langle \hat{j}_{x}(\textbf{q},\tau) \langle \hat{j}_{x}(\textbf{-q},0) \rangle$, and $\hat{j}_{x}(\textbf{q},\tau)$ is the $(\textbf{q},\tau)$-dependent current operator in $x$ direction.
The validity of Eq.\ref{2} has been  proved for metal-insulator transitions in the Hubbard model in many works\cite{PhysRevLett.75.312,PhysRevLett.83.4610,PhysRevLett.120.116601}.
We also define $N(0)$, the density of states at the Fermi level as $N(0) \simeq \beta \times G (\textbf{r}=0,\tau=\beta/2)$ \cite{PhysRevLett.75.312} to differentiate metal phase and insulating phase, where $G$ is the imaginary-time dependent Greens function.

With the aim of exploring the system evolving with the variation in the magnetic order, we computed the antiferromagnetic (AFM) spin structure factor,
\begin{equation}
S_{AFM} = \frac{1}{N_s}\langle [\sum_{lr}(\hat{S}_{lar}^z-\hat{S}_{lbr}^z)]^2 \rangle,
\label{3}
\end{equation}
which indicates the onset of long-range AFM order if $\lim_{N_s \to \infty}(S_{AFM}/{N_s})>0$.
Here, $N_s$ represents the number of lattice sites, $\hat{S}_{lar}^z(\hat{S}_{lbr}^z)$ is the $z$-component spin operator on the A (B) sublattice of layer $l$.
$S_{AFM}$ for different interactions are calculated on lattices with $L = 6, 9, 12, 15$ and $18$ are extrapolated to the thermodynamic limit using polynomial functions in $1/\sqrt{N_s}$.

\begin{figure}[tbp]
\includegraphics[scale=0.4]{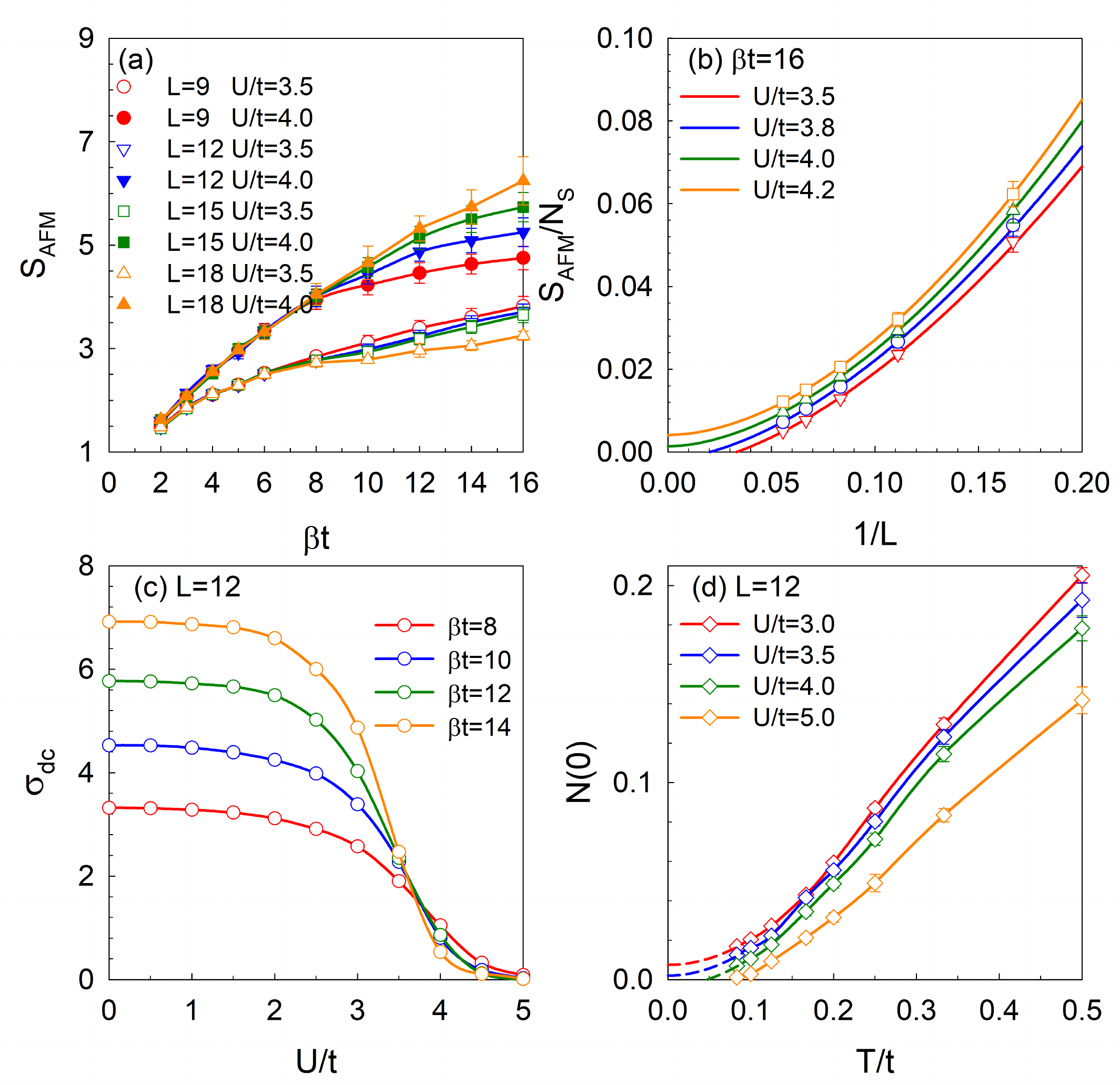}
\caption{(a) The AFM spin structure factor $S_{AFM}$ depends on $\beta = 1/T$ with different  lattice size and interaction strength; (b) scaling behavior of the normalized AFM spin structure factor $S_{AFM}/N_S$ for corresponding $U$ values at $\beta t = 16$.
Solid lines represent third-order polynomial fits to the data in $1/\sqrt{N_S}$.
(c) the conductivity $\sigma_{dc}$ at half-filling versus the interaction U for different temperatures;
(d) density of states at the Fermi energy $N(0)$ versus the temperature T for various U's.
With U increasing, $N(0)$ gradually decreases, and when $N(0)$ tends to zero at $T/t \to 0$, the system transitions from a metallic phase to a Mott insulator.}
\label{S_AFM}
\end{figure}

\noindent
\underline{\it Results and discussion}
When the electronic properties are concerned, values of $T/t \textless 0.5$ or so have been found sufficiently low for strong electron correlations to manifest themselves\cite{PhysRevB.86.245117,PhysRevB.99.205434}.
Specifically in the Hubbard model, the low- and high-temperature regions could be determined by behaviors of
spin fluctuations and charge fluctuations, whose boundary is near $T/t \textless 1.0$\cite{PhysRevB.63.125116}.
In order to address ground-state properties by use the finite temperature quantum Monte Carlo method, we carefully monitor our simulations and make sure the results are converged at low enough temperatures.
As plot in Fig.\ref{S_AFM}(a), the antiferromagnetic spin structure factor $S_{AFM}$ as a function of inverse temperature $\beta = 1/T$ acquired on different lattice sizes $L$ and interaction strength $U$.
The AFM order increases as the temperature is lowered when $T$ drops below a lattice-dependent temperature, $S_{AFM}$ saturates and has very little $\beta$ dependence within statistical errors.
So we reasonably conclude that the physical observable has reached the $T/t = 0$ ground state if its value is convergent below some $\beta_0 t \sim 10$, which is consistent with the previous findings\cite{PhysRevLett.120.116601,PhysRevB.101.245161}.
In the following results, the lowest temperature has been reached, at least, $T/t = 0.1$, which is sufficiently low to investigate physics properties of the low-temperature region.
To explore the effect of interlayer coupling, we fix $t_c=0.138t$ here and $t_c$ will be varied in the following.
From Fig.\ref{S_AFM}(a), one can see that $S_{AFM}$ is almost independent with $L$ in all zone at $U/t=3.5$, whereas at $U/t=4.0$, $S_{AFM}$ increases significantly with increasing lattice size $L$ for $\beta t > 8$, indicating a possibility of a long-range order at $U/t \sim 4.0$.
Figure \ref{S_AFM}(b) presents the finite-size scaling results of the AFM spin structure factor $S_{AFM}/{N_s}$.
By extrapolating the data to the thermodynamic limit, we estimate the critical point of the AFM long-range order to be $U_c/t \sim 4.0$, which is similar to the previous findings\cite{PhysRevLett.120.116601,HUANG2019310,PhysRevB.101.155413}.

To reveal a more interesting electronic property as Mott-like insulating behavior in ABC-TLG, we present Fig.\ref{S_AFM}(c) to show the conductivity $\sigma_{dc}$ as a function of interaction $U$ at half-filling for different temperatures.
The conductivity $\sigma_{dc}$ monotonically decreases with increasing $U$, for the same $U$, $\sigma_{dc}$ values at higher temperature exceeds those of lower temperature.
The intersection of the curves defines the critical field $U_c$, representing the transitions from metal to Mott insulator, which emerges within the range of $3.5 \textless U_c/t \textless 4.0$.
To further confirm the transition, we calculated the density of states at the Fermi level $N(0)$ around the transition driven by $U$ in Fig.\ref{S_AFM}(d).
The interaction-induced Mott insulator, characterized by the opening of a Mott gap, results that $N(0)$ tends to 0 when $T/t \to 0$ \cite{PhysRevLett.117.146601}, which is observed in the range of $3.5 \textless U_c/t \textless 4.0$.
A finite y-axis intercept in the $T/t \to 0$ limit indicates that the metallic phase exists.
Since conditions at even lower temperatures are challenging, our polynomial fittings at $T/t \to 0$ (dashed lines) are to be interpreted on a qualitative level.

\begin{figure}[tbp]
\includegraphics[scale=0.4]{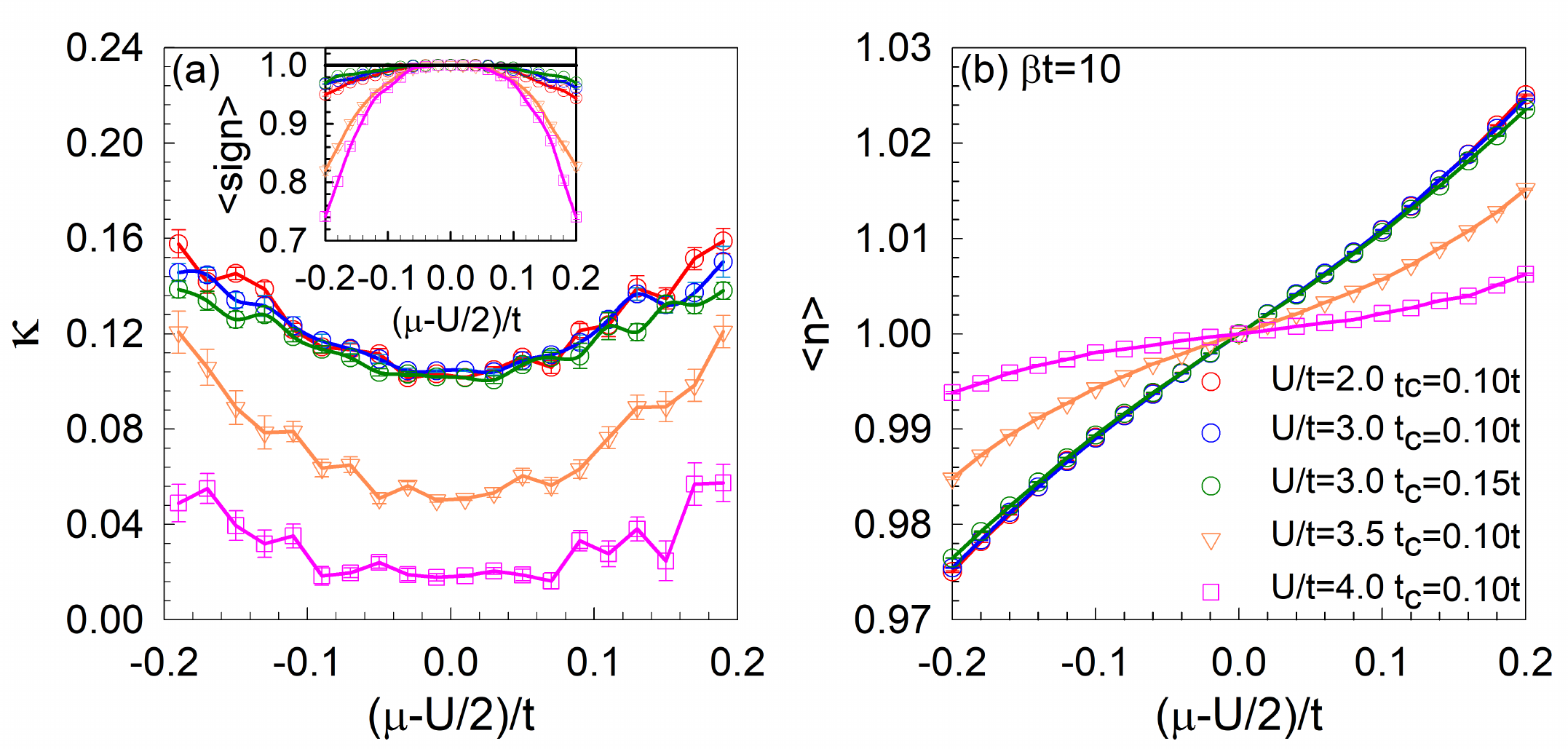}
\caption{(a) Charge compressibility $\kappa$ and (b) electron filling $\langle{n}\rangle$ versus $\mu$ at $\beta t=10$ for several interaction
strengths and interlayer hopping energy. the inset: the corresponding $\langle{sign}\rangle$ for different values of U and $t_c$ at $\beta t=10$.
} \label{Kappa}
\end{figure}

The half-filled Hubbard model on a honeycomb lattice exhibits a charge(Mott) excitation gap at a sufficiently large $U$\cite{PhysRevLett.120.116601,PhysRevX.6.011029}.
On the other hand, the non interacting Anderson insulator is gapless at the Fermi level\cite{RevModPhys.50.191,IJOMPB24}.
So the gap can also be used to establish the existence of the Mott insulator even although there is no association between the gap and the symmetry breaking.
Basically the single-particle gap can be extracted from the density of states, and here we deduce the energy gap information by examining the behavior of charge compressibility $\kappa(\mu)=d{\langle\hat{n}(\mu)\rangle}/d(\mu)$ at the Fermi level, where $\langle\hat{n}(\mu)\rangle$ is the average density at chemical potential $\mu$.
Results for $\kappa(\mu)$ evaluated at inverse temperature $\beta t = 10$ are depicted in Fig.\ref{Kappa} for $L=9$ with various $t_c 's$ and $U's$.
In the thermodynamic limit, $\kappa$ of a system with an energy gap will disappear at $T/t=0$.
However, due to the temperature broadening effect, the threshold of $\kappa$ is finite on finite lattices at non zero temperature, and $\kappa=0$ as a criterion will overestimate the critical coupling strength\cite{PhysRevB.76.144513}.
Therefore, we take $\kappa \sim 0.04$ as an appropriate threshold to distinguish between gapped and gapless phases analyzing the effect of the finite $T$ and the non interaction limit\cite{PhysRevLett.120.116601,HUANG2019310}.

Suggested from Fig.\ref{Kappa}(a), for $t_c=0.10t$, the system becomes incompressible at $U_c/t \simeq 3.5 \sim 4.0$, combining results shown in Fig.\ref{S_AFM}, we identify that the state at half filling with $U > U_c$ is an antiferromagneticlly ordered Mott insulating state.
Moreover, $\kappa(\mu)$ is insensitive with the change in interlayer coupling strength.
From Fig.\ref{Kappa}(b), we can see that $\langle{n(\mu)}\rangle$ converges faster than $\kappa$ vanishes.
Tuning $\mu$ away from half-filling breaks the particle-hole symmetry and leads to a sign problem.
For the present results, our numerical results are reliable, in the inset of Fig.\ref{Kappa}(a), one can see that $\langle{sign}\rangle$ is mostly larger than 0.74 for $\kappa$ at $\beta t = 10$ with $t_c=0.10t,0.15t,0.20t$ and $U/t \leq 4.0$.
Comparing to the results of monolayer and bilayer graphene in previous studies\cite{PhysRevLett.120.116601,HUANG2019310}, one may argue that the location of critical point $U_c$ where the gap opens in graphene systems is mostly dominated by the hexagonal lattice structure.

To investigate the superconducting property of ABC-TLG, we studied the effective pairing interaction with different pairing symmetries.
Following previous studies\cite{PhysRevB.39.839,PhysRevLett.110.107002,PhysRevB.40.506}, pairing susceptibility is defined as
\begin{equation}
P_\alpha = {\frac{1}{N_S}}{\sum_{lij}\int_{0}^{\beta}{\mathrm{d}\tau}\langle \Delta_{l\alpha}^\dagger(i,\tau)\Delta_{l\alpha}(j,0)\rangle},
\label{4}
\end{equation}
where $\alpha$ stands for the pairing symmetry.
Due to the constraint of on-site Hubbard interaction in Eq.(\ref{1}), the corresponding order parameter $\Delta_{l\alpha}^\dagger(i)$ reads
\begin{equation}
\Delta_{l\alpha}^{\dagger}(i) = \sum_{\textbf{l}}f_{\alpha}^{\dagger}(\delta_{\textbf{l}})(a_{li\uparrow}b_{li+\delta_{\textbf{l}}\downarrow}-a_{li\downarrow}b_{li+\delta_{\textbf{l}}\uparrow})^{\dagger},
\label{5}
\end{equation}
with $f_\alpha(\delta_\textbf{l})$ being the form factor of the pairing function.
Here, in Eq.\ref{5}, the vectors $\delta_\textbf{l} (\textbf{l} = 1-3)$ denote the nearest neighbor (NN) intersublattice connections, as sketched in Fig.\ref{structure}(b) of Ref. \cite{PhysRevB.84.121410}.
Considering that the symmetry of the honeycomb lattice is governed by the $D6$ point group, three possible NN pairing symmetries are characterized by: (a) extended $S (ES)$, (b) $d + id$, and (c) the $p + ip$ wave\cite{PhysRevB.84.121410,PhysRevB.90.245114,HUANG2019310,Ma2015EPL}.
These extended pairing symmetries are defined with different phase shifts upon $\pi/3$ or $2\pi/3$ rotations.
The form factors of the singlet ES wave and NN-bond $d + id$ pairing are given by
\begin{equation}
f_{ES}(\delta_\textbf{l}) = 1, \textbf{l} = 1-3,
\label{6}
\end{equation}
\begin{equation}
f_{d+id}(\delta_\textbf{l}) = e^{i(\textbf{l}-1){\frac{2\pi}{3}}}, \textbf{l} = 1-3,
\label{7}
\end{equation}
as for the NN-bond $f_{p+ip}$ pairings, the form factor of A and B sublattices are different, where
\begin{equation}
f_{p+ip}(\delta_\textbf{al}) = e^{i(\textbf{l}-1){\frac{2\pi}{3}}}, \textbf{l} = 1-3,
\label{8}
\end{equation}
\begin{equation}
f_{p+ip}(\delta_\textbf{bl}) = e^{i[(\textbf{l}-1){\frac{2\pi}{3}}+\pi]}, \textbf{l} = 1-3.
\label{9}
\end{equation}
for A and B respectively, which are pretty similar except that there is a $\pi$ phase shift.
In addition to the NN-bond pairings, we also studied longer-range pairings by the adding next-nearest-neighbor bond pairing for $d + id$ wave symmetry, which have the following form factors:
\begin{equation}
f_{d+id}(\delta_\textbf{l}) = e^{i(\textbf{l}-1){\frac{2\pi}{3}}}, \textbf{l} = 1,2,3...6.
\label{10}
\end{equation}

$P_\alpha$ includes both the renormalization of the propagation of the individual particles and the interaction vertex between them, whereas $\widetilde{P}_\alpha$ includes only the former effect. In order to extract the effective pairing interaction in a finite system, one should subtract from $P_\alpha$ its uncorrelated single-particle contribution $\widetilde{P}_\alpha$, which is achieved by replacing $\langle a_{li\downarrow}^{\dagger}a_{lj\downarrow}b_{i+\delta_\textbf{l}\uparrow}^{\dagger}b_{j+\delta_{\textbf{l}^{'}}\uparrow}\rangle$ in Eq.(\ref{4}) with $\langle a_{i\downarrow}^{\dagger}a_{j\downarrow}\rangle \langle b_{i+\delta_\textbf{l}\uparrow}^{\dagger}b_{j+\delta_{\textbf{l}^{'}}\uparrow}\rangle$, and the effective pairing interaction $\textbf{P}_\alpha$ is defined as $\textbf{P}_\alpha=P_\alpha-\widetilde{P}_\alpha$.

Distinguished $P_\alpha$ and $\widetilde{P}_\alpha$, we are allowed to extract the interaction vertex $\Gamma_\alpha$:
\begin{equation}
\Gamma_\alpha = \frac{1}{P_\alpha}-\frac{1}{\widetilde{P}_\alpha}.
\label{11}
\end{equation}
If $\Gamma_\alpha\widetilde{P}_\alpha<0$, the associated pairing interaction is attractive.
In fact, Eq.(\ref{4}) can be rewritten as
\begin{equation}
P_\alpha = \frac{\widetilde{P}_\alpha}{1+\Gamma_\alpha\widetilde{P}_\alpha}
\label{12}
\end{equation}
suggests that $\Gamma_\alpha\widetilde{P}_\alpha \to -1$ signals a superconducting instability.
This is the analog of the familiar Stoner criterion $U\chi_0=1$, which arises from the random phase approximation expression $\chi=\chi_0/(1-U\chi_0)$ for the interacting magnetic susceptibility $\chi$ in terms of the noninteracting $\chi_0$.

\begin{figure}[tbp]
\includegraphics[scale=0.4]{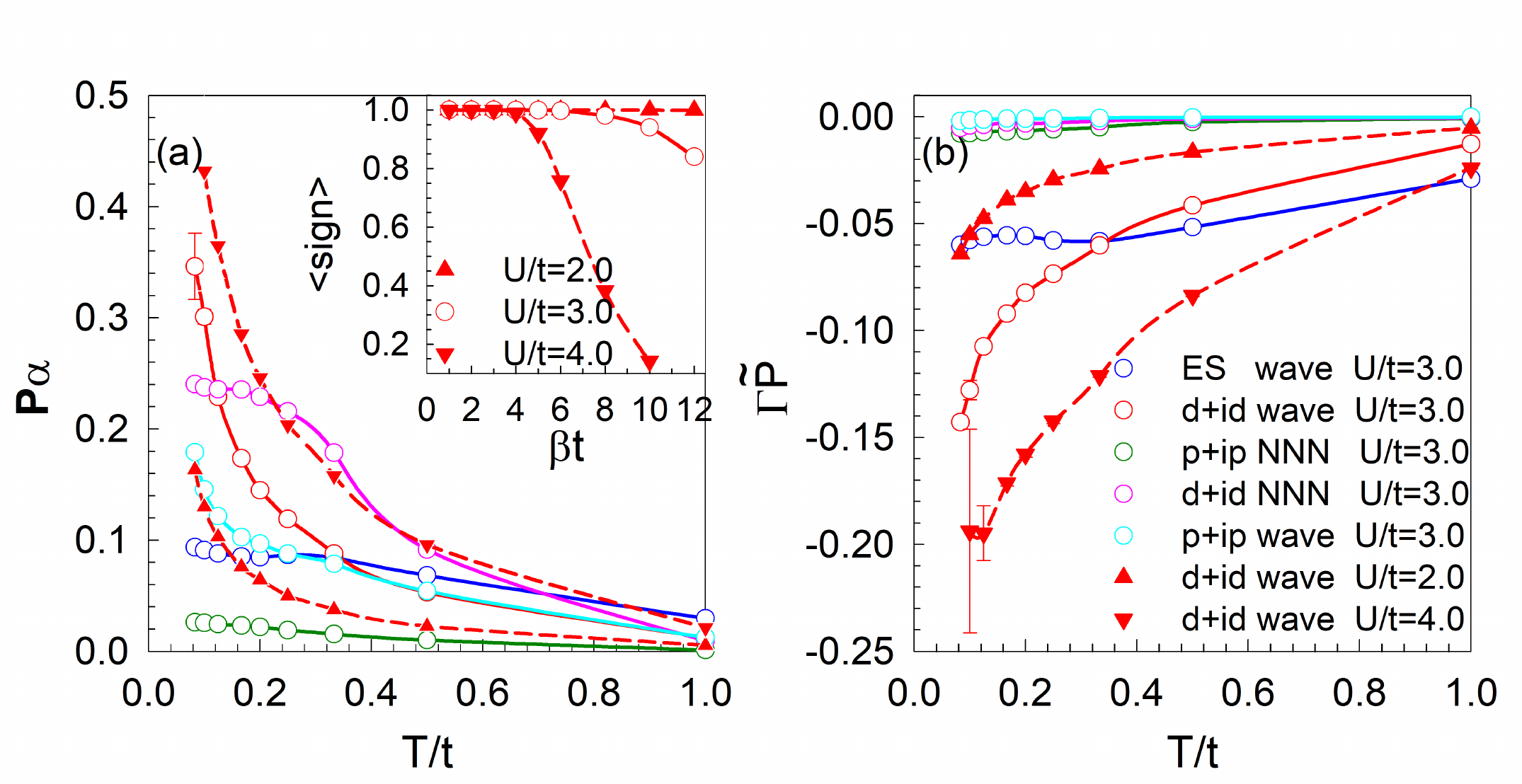}
\caption{(a) Effective pairing interaction $\textbf{P}_\alpha$ of different pairing symmetries and (b) product of superconducting vertex $\Gamma$ and no-vertex pairing susceptibility $\widetilde{P}_\alpha$ as a function of temperature.
Parameters are electron filling $\langle n\rangle=0.95$ and interlayer coupling $t_c$=0.10$t$.
If $\Gamma\widetilde{P}\to-1$, a superconducting instability ensues.
The inset: the temperature-dependent $\langle sign\rangle$ at $\langle n\rangle=0.95$ with the corresponding $U$ for $t_c$=0.10$t$.
} \label{different U}
\end{figure}

Figure \ref{different U}(a) shows the temperature dependence of $\textbf{P}_\alpha$ for different pairing symmetries at $\langle n\rangle=0.95$ with $t_c=0.10t$.
The effective pairing interaction for various symmetries increase as the temperature is lowered and, most remarkably, the $d+id$ pairing symmetry dominates other symmetries at relatively low temperatures.
The effective pairing susceptibility $\textbf{P}_{d+id}$ with $U/t = 2.0$ and $U/t = 4.0$ are also shown, in comparison with $U/t = 3.0$ from which one can see that the $d+id$ pairing interaction is enhanced greatly as the value of $U$ increases.

In Fig.\ref{different U}(b), we examine the effect of the interaction vertex $\Gamma_\alpha\widetilde{P}_\alpha$, signaling a superconducting instability by $\Gamma_\alpha\widetilde{P}_\alpha \to -1$.
The tendency to pairing becomes greater as the temperature is lowered, especially to the $d+id$ wave where the effect of temperature on $\Gamma\widetilde{P}$ is more pronounced than other pairing symmetries.
One also can see that the growth in pairing vertex from interaction strengths $U/t = 2.0$ to $U/t = 4.0$, which is consistent with Fig.\ref{different U}(a) that $d+id$ symmetry is significantly dependent on $U$.

\begin{figure}[tbp]
\includegraphics[scale=0.4]{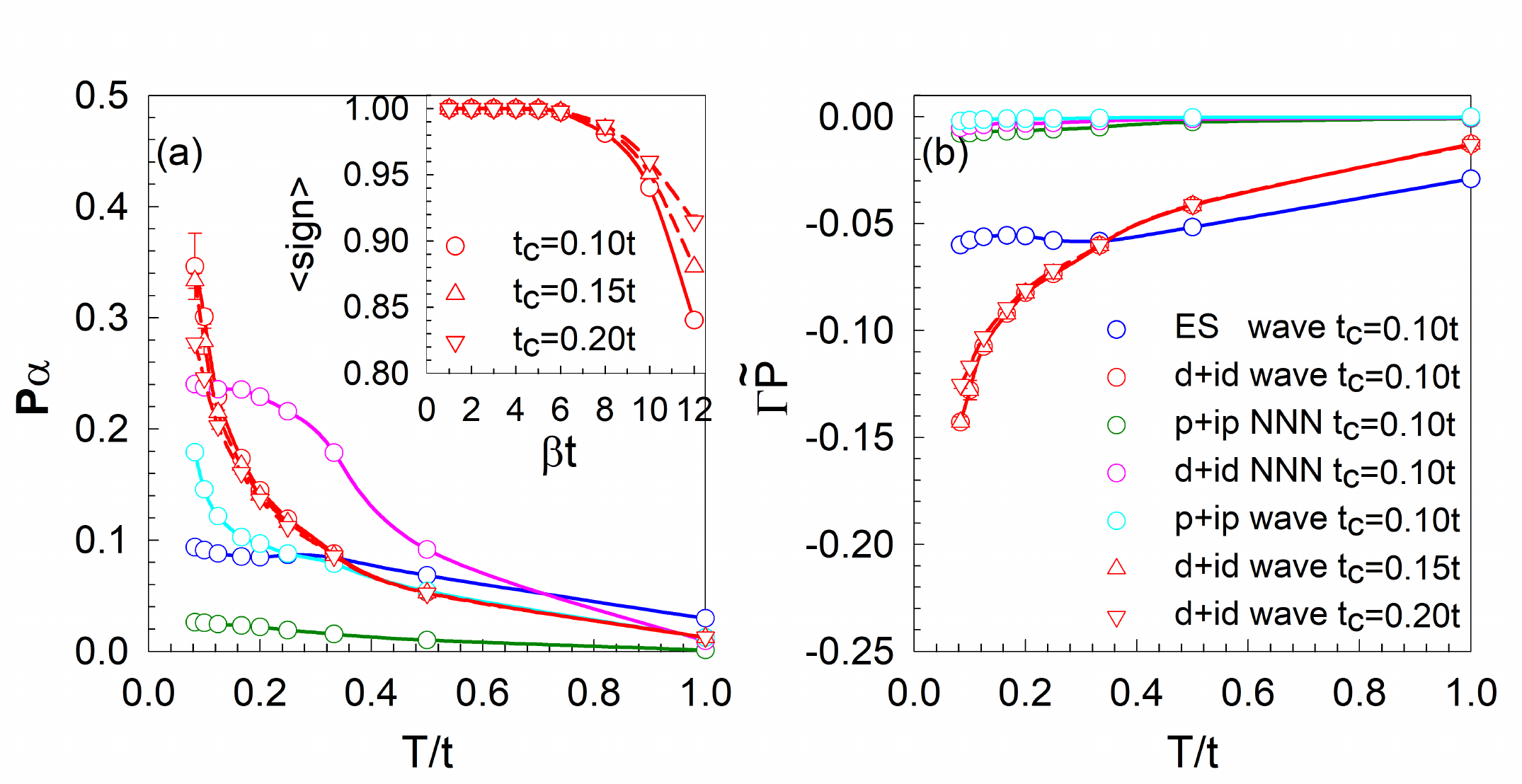}
\caption{The tendency to (a) effective pairing interaction $\textbf{P}_\alpha$ and (b) superconductivity $\Gamma\widetilde{P}$ as functions of temperature at electron filling $\langle n\rangle = 0.95$ and interaction strength $U/t=3.0$ for different interlayer couplings $t_c$.
The inset: the temperature-dependent $\langle sign\rangle$ at $\langle n\rangle = 0.95$ with the corresponding $t_c$ for $U/t=3.0$.
} \label{different t_c}
\end{figure}

We also studied the temperature dependence of effective pairing interaction $\textbf{P}_\alpha$ and superconducting instability $\Gamma\widetilde{P}$ with different interlayer couplings $t_c$ in Fig.\ref{different t_c}.
One can see that both the effective pairing interaction and the superconducting instability of $d+id$ symmetry are almost independent with the interlayer coupling strength.
Suggested from Figs.\ref{different U} and \ref{different t_c}, the positive $\textbf{P}_\alpha$ indicates that there actually generate effective attractions between electrons in the system at low temperatures.
This also demonstrates that the electron-electron correlation plays a key role in driving the superconductivity.
In addition, $\langle sign\rangle$ is larger than 0.83 for $t_c=0.10t, 0.15t, 0.20t$ at $U/t = 3.0$ as shown in the inset of Fig.\ref{different t_c} (a) and larger than 0.80 for $t_c=0.10t$ at $U/t < 4.0$ shown in the insert of Fig.\ref{different U} (a).
At a larger $U/t = 4.0$, $\langle sign\rangle$ is mostly larger than 0.35 as $\beta t \leq 8$.
For the cases of $U/t = 4.0$ and $\beta t > 8$, the sign problem is worse which is not important as the dominant pairing symmetry is robust on the temperature.

\noindent
\underline{\it Conclusions}---
To summarize, we perform a quantum Monte Carlo study of the charge compressibility, spin correlation and superconducting instability in the ABC graphene trilayer system.
The results of the charge compressibility and spin correlation show that, at half-filling, an antiferromagnetically ordered Mott insulator is proposed beyond a critical $U_c/t \sim 4.0$. With finite doping, the superconducting pairing with $d+id$ symmetry dominates over other pairing symmetries.
We also analyze the effect of the on-site interaction and the interlayer interaction in superconductivity.
It is found that the dominant $d+id$ superconducting pairing interaction increases with increasing on-site interaction strength,
which means that the $d+id$ superconductivity is driven by the strong on-site interaction.
The results presented here demonstrate the interaction-driven superconductivity with a dominant $d+id$ pairing symmetry in ABC-TLG,
and the superconductivity is arising from a doped Mott insulator.

\noindent
\underline{\it Acknowledgment}---
This work was supported by NSFC (Grants No. 11774033 and No. 11974049) and Beijing Natural Science Foundation (Grant No. 1192011).
The numerical simulations in this work were performed at HSCC of Beijing Normal University and Tianhe in the Beijing Computational Science Research Center.
\bibliography{ref}
\end{document}